\documentclass[sigconf]{acmart}
\usepackage{tabularx}
\usepackage{tabularray}
\AtBeginDocument{%
  }

\copyrightyear{2025}
\acmYear{2025}
\setcopyright{acmlicensed}\acmConference[MM '25]{Proceedings of the 33rd ACM International Conference on Multimedia}{October 27--31, 2025}{Dublin, Ireland}
\acmBooktitle{Proceedings of the 33rd ACM International Conference on Multimedia (MM '25), October 27--31, 2025, Dublin, Ireland}
\acmDOI{10.1145/3746027.3755292}
\acmISBN{979-8-4007-2035-2/2025/10}

\begin{document}

%%
%% The "title" command has an optional parameter,
%% allowing the author to define a "short title" to be used in page headers.
\title{Joint Holistic and Lesion Controllable Mammogram Synthesis via Gated Conditional Diffusion Model}
\renewcommand{\shorttitle}{Mammogram Synthesis via Gated Conditional Diffusion Model}

%%
%% The "author" command and its associated commands are used to define
%% the authors and their affiliations.
%% Of note is the shared affiliation of the first two authors, and the
%% "authornote" and "authornotemark" commands
%% used to denote shared contribution to the research.

\author{Xin Li}
\authornotemark[1]
\affiliation{%
\institution{Wuhan National Laboratory for Optoelectronics, Huazhong University of Science and Technology}
\city{Wuhan}
\country{China}}
\email{lixin2023@hust.edu.cn}
\orcid{https://orcid.org/0009-0004-7815-3426}

\author{Kaixiang Yang}
\authornote{Co-first authors.}
\affiliation{%v
\institution{Wuhan National Laboratory for Optoelectronics, Huazhong University of Science and 
    Technology}
\city{Wuhan}
\country{China}}
\email{kxyang@hust.edu.cn}
\orcid{https://orcid.org/0009-0008-9160-8498}

\author{Qiang Li}
\affiliation{%
\institution{Wuhan National Laboratory for Optoelectronics, Huazhong University of Science and 
    Technology}
\city{Wuhan}
\country{China}}
 \email{liqiang8@hust.edu.cn}
 \orcid{https://orcid.org/0000-0002-9815-4432}

\author{Zhiwei Wang}
\authornote{Corresponding author.}
\affiliation{%
\institution{Wuhan National Laboratory for Optoelectronics, Huazhong University of Science and 
    Technology}
\city{Wuhan}
\country{China}}
\email{zwwang@hust.edu.cn}
\orcid{https://orcid.org/0000-0002-1612-8573}

%%
%% By default, the full list of authors will be used in the page
%% headers. Often, this list is too long, and will overlap
%% other information printed in the page headers. This command allows
%% the author to define a more concise list
%% of authors' names for this purpose.
\renewcommand{\shortauthors}{Xin Li, Kaixiang Yang, Qiang Li, Zhiwei Wang}

%%
%% The abstract is a short summary of the work to be presented in the
%% article.
\begin{abstract}
Mammography is the most commonly used imaging modality for breast cancer screening, driving an increasing demand for deep-learning techniques to support large-scale analysis. However, the development of accurate and robust methods is often limited by insufficient data availability and a lack of diversity in lesion characteristics.
While generative models offer a promising solution for data synthesis, current approaches often fail to adequately emphasize lesion-specific features and their relationships with surrounding tissues.
In this paper, we propose Gated Conditional Diffusion Model (GCDM), a novel framework designed to jointly synthesize holistic mammogram images and localized lesions. 
GCDM is built upon a latent denoising diffusion framework, where the noised latent image is concatenated with a soft mask embedding that represents breast, lesion, and their transitional regions, ensuring anatomical coherence between them during the denoising process.
To further emphasize lesion-specific features, GCDM incorporates a gated conditioning branch that guides the denoising process by dynamically selecting and fusing the most relevant radiomic and geometric properties of lesions, effectively capturing their interplay.
Experimental results demonstrate that GCDM achieves precise control over small lesion areas while enhancing the realism and diversity of synthesized mammograms. These advancements position GCDM as a promising tool for clinical applications in mammogram synthesis. Our code is available at \url{https://github.com/lixinHUST/Gated-Conditional-Diffusion-Model/}
\end{abstract}

%%
%% The code below is generated by the tool at http://dl.acm.org/ccs.cfm.
%% Please copy and paste the code instead of the example below.
%%


%%
%% Keywords. The author(s) should pick words that accurately describe
%% the work being presented. Separate the keywords with commas.
%% A "teaser" image appears between the author and affiliation
%% information and the body of the document, and typically spans the
%% page.

% \received{20 February 2007}
% \received[revised]{12 March 2009}
% \received[accepted]{5 June 2009}

%%
%% This command processes the author and affiliation and title
%% information and builds the first part of the formatted document.
\maketitle
\section{INTRODUCTION}
Breast cancer, the most prevalent malignant tumor among middle-aged and elderly women, impacts approximately 1.2 million women annually~\cite{hosseini2016early}. Traditional immunohistochemical analysis is invasive, complex, and often insufficient for precise diagnosis and personalized treatment. In contrast, radiomics-based breast cancer diagnosis~\cite{lu2018machine} provides a non-invasive, comprehensive, and cost-effective alternative. 
Among imaging modalities, mammography has demonstrated significant efficacy in early detection and diagnosis~\cite{moss2012impact}. Recently, advancements in deep learning have led to the development of computer-aided diagnosis (CAD) systems that leverage mammography to improve large-scale screening, enabling faster and more objective clinical decision-making.

However, deep learning-based CAD models typically require large volumes of high-quality training data, and acquiring accurately annotated medical images is both costly and time-intensive~\cite{wang2021annotation,wang2023pymic}. Generative models present a promising solution by simulating the data distribution of real images to synthesize realistic samples, thereby facilitating efficient and cost-effective data augmentation. 
Among these, conditional image synthesis provides a more targeted approach compared to unconditional generation. By leveraging conditional generation, it becomes possible to better control the class and spatial diversity of synthesized images, thereby addressing specific data deficiencies in a focused manner.

Currently, conditional image synthesis is primarily achieved through two approaches: generative adversarial networks (GANs) and denoising diffusion probabilistic models (DDPMs). 
GAN-based methods, such as Pix2Pix~\cite{pix2pix}, have demonstrated remarkable success in high-quality image synthesis. These methods~\cite{StyleGAN2,pSp,pix2pixhd,SPADE,karras2019style} typically utilize a UNet generator to transform input masks into realistic images and a PatchGAN discriminator to differentiate between real image-mask pairs and generated ones. Despite their achievements, GANs are plagued by inherent limitations, including unstable adversarial training dynamics~\cite{gansurvey}, which can compromise their reliability and overall performance.
In contrast, DDPM-based methods have emerged as a robust alternative, generating images by iteratively denoising Gaussian noise through a learned reverse process~\cite{DDPM}. This approach, explored in various studies~\cite{xreal,Controlnet,Segguidediff,semanticDM,CDM,DiffusionBeatGan}, offers more stable training and enhanced synthesis quality~\cite{chen2024medical}, making it a promising avenue for advancing conditional image synthesis.

Based on the aforementioned analysis, most current methods for mammogram synthesis are rooted in DDPM-based approaches, which integrate control conditions during the denoising process to guide the generation of mammograms with desired lesions. For example, Seg-Diff~\cite{Segguidediff} leverages semantic segmentation masks as conditioning inputs to direct the diffusion model toward generating anatomically coherent structures. However, these methods predominantly emphasize global control and often fall short in achieving precise fine-grained lesion control.
Mammogram synthesis, as a specialized task in medical image generation, places significant emphasis on lesion-controllable synthesis to ensure both anatomical fidelity and clinical relevance. Beyond producing visually realistic mammograms, it is essential to preserve the intricate relationships between lesions and their surrounding tissues, as well as to maintain clinically meaningful attributes that are critical for accurate diagnosis and analysis. 

In this paper, we present the Gated Conditional Diffusion Model (GCDM), a novel framework designed to enhance the quality and diversity of mammogram synthesis while achieving precise control over local lesion regions. Built upon a fine-tuned image-conditioned stable diffusion model, GCDM integrates the noised latent image with the embedding of a soft mask label through channel concatenation, ensuring anatomical coherence between lesions and surrounding tissues. The soft mask label is generated by applying a blurring operation to the hard label, facilitating smooth transitions at lesion boundaries.

To further strengthen lesion control, we introduce a gated conditioning branch that dynamically selects and fuses the most relevant radiomic and geometric properties of lesions. Leveraging a gate fusion network with a top-$k$ strategy, this branch maximizes the correlation between lesion features and provides high-quality mass feature representations for mammogram synthesis.

Our key contributions are summarized as follows:
\begin{itemize}
\item We propose the GCDM framework, which enables holistic mammogram synthesis with precise lesion control. By channel-wise concatenation of the soft mask label with the noisy latent image in the latent diffusion model, our approach ensures global anatomical consistency between lesions and surrounding tissues.
\item We introduce a gated conditioning branch that dynamically selects and fuses the most relevant radiomic and geometric features of lesions using a top-$k$ strategy. This innovation preserves diagnostic consistency and enhances lesion control during the synthesis process.
\item We conduct a comprehensive quantitative evaluation of GCDM on the VinDr-Mammo dataset~\cite{dataset}. Experimental results in Section~\ref{AA} demonstrate that our method outperforms state-of-the-art approaches in mammogram synthesis quality (FID), global anatomical coherence (PA, Breast IoU), and precise lesion control (Mass IoU).
\end{itemize}

\section{RELATED WORKS} \label{related}

\subsection{GAN-based Controllable Image Synthesis}
Generative Adversarial Networks (GANs)~\cite{GAN} consist of a generator and a discriminator, where the generator maps a low-dimensional latent space to high-dimensional image space, while the discriminator distinguishes real from generated images, providing feedback to improve synthesis quality. 
In controllable image synthesis, GANs are trained with adversarial and reconstruction losses, enabling mask-to-image translation. The generator learns to produce images that align with input masks, while the discriminator differentiates real image-mask pairs and generated image-mask pairs. Methods like Pix2pix~\cite{pix2pix} employ a UNet generator and a PatchGAN discriminator to refine mask-to-image translation, achieving remarkable success in high-quality image synthesis, while Pix2pixHD~\cite{pix2pixhd} enhances resolution with a multi-scale generator. SPADE~\cite{SPADE} improves semantic alignment using spatially-adaptive normalization in generator.

Unlike the aforementioned methods that primarily rely on adversarial loss between the generator and discriminator, the pSp~\cite{pSp} framework is a GAN inversion method, provides a fast and accurate solution for encoding masks into the $W+$ latent space of a pretrained StyleGAN~\cite{karras2019style} generator. It employs an FPN-based encoder to extract style vectors at different pyramid scales, which are then injected into the generator at corresponding spatial levels. The optimization process incorporates pixel-wise reconstruction loss, perceptual loss, and average latent code regularization to maintain structural coherence.

Although GAN-based methods have achieved remarkable success, they inherently suffer from instability in adversarial training, which can result in mode collapse and challenges in convergence~\cite{gansurvey}. 

\subsection{DDPM-based Controllable Image Synthesis}
Diffusion models have emerged as a powerful approach for image generation, progressively denoising a random noise input over multiple steps. At the core of this process is the UNet denoising network, which predicts the noise in a noisy image based on the current time step. During inference, images can be generated using the DDPM strategy~\cite{DDPM}, or alternatively, the DDIM strategy~\cite{DDIM}, which reduces the number of denoising steps while maintaining high-quality outputs, enabling faster image generation.

DDPM-based semantic image synthesis involves conditioning each denoising step on semantic masks, enabling the generation of semantically coherent images. Currently, there are two primary classes of methods for incorporating conditioning inputs into the denoising process: concatenation-based and cross-attention-based approaches.

The concatenation-based approach directly integrates conditioning features with the noisy image through channel concatenation. For example, SR3~\cite{SR3} concatenates low-resolution images with noisy images to guide the denoising process, producing high-resolution outputs. Similarly, Zero123~\cite{zero123} employs the reference view image as a control condition, concatenating it with the noisy image during denoising to generate the target view image. This method provides a straightforward and efficient way to achieve global control over the generated images. Beyond concatenating with noisy inputs, some methods integrate conditioning features with intermediate features in the UNet through skip connections. For instance, ControlNet~\cite{Controlnet} duplicates the UNet’s downsampling layers to create a control encoder, ensuring that conditioning features are processed at corresponding network depths before being injected back via zero convolutions for precise guidance. T2I-adapter~\cite{mou2024t2i} follows a similar approach but uses an external lightweight network to extract conditioning features, aligning them with UNet features at different layers to enhance controllability while preserving the pretrained model's structure.

The cross-attention-based approach, on the other hand, leverages a cross-attention mechanism~\cite{vaswani2017attention} to enable deeper interaction between conditioning features and the denoising process. This approach is most commonly used with text as the conditioning input. In this architecture, the $Q$ matrix is derived from UNet features, while the $K$ and $V$ matrices are derived from the embedding of conditioning inputs. Stable diffusion models~\cite{CDM} utilize this mechanism to interact with text features through a cross-attention layer, achieving fine-grained control over generated images. Extending this, the IP-adapter~\cite{ye2023ip} introduces style image encoding features, enabling precise control over image style through decoupled cross-attention in addition to text-based guidance. This approach excels in fine-grained control, allowing for precise manipulation of localized regions, such as lesions, while maintaining anatomical consistency.

In summary, concatenation-based methods offer simplicity and efficiency for global control, while cross-attention-based methods provide superior fine-grained control, making them particularly suitable for tasks requiring detailed manipulation of specific image regions.

\begin{figure*}[ht]
\centering
\includegraphics[width=\linewidth]{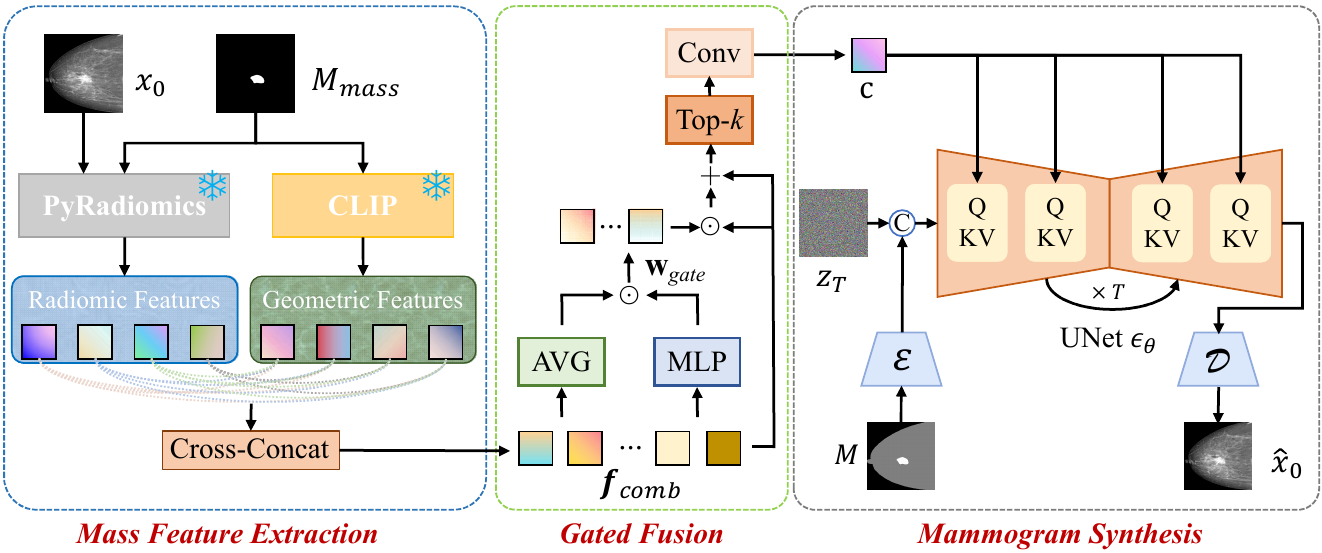}
\caption{An overview of the proposed Gated Conditional Diffusion Model (GCDM) for precise and effective mammogram synthesis, ensuring global anatomical consistency between lesions and surrounding tissues, while preserving diagnostic accuracy and enhancing lesion controllability. } \label{method}
\end{figure*}

\subsection{Medical Image Synthesis}
Medical image generation, particularly mammogram synthesis, requires a high degree of anatomical accuracy and precise lesion control. GAN-based methods, such as VTGAN~\cite{kamran2021vtgan}, which employ a coarse-to-fine generator and a Vision Transformer-based discriminator for angiogram synthesis, often suffer from training instability. Prior-Guided GAN~\cite{priorgan} incorporates tissue priors in class-conditional GAN, such as density and lesion maps. By fusing these priors with latent codes, it generates realistic, class-specific mammograms that preserve anatomical details. 

In contrast, DDPM-based models, such as Seg-Diff~\cite{Segguidediff}, generate breast MRI from anatomical segmentation by concatenating the mask with the noisy image in the image space to guide the denoising process. MAM-E~\cite{montoya2024mam} and RadiomicsFill~\cite{radiomicsfill} both utilize conditional diffusion models for mammogram tumor inpainting, leveraging prompt text and radiomics features, respectively. Compared to GANs, diffusion models exhibit greater training stability and are less prone to mode collapse, offering a broader diversity of output samples~\cite{DiffusionBeatGan}. These models provide a more reliable framework for incorporating semantic masks, thereby improving anatomical alignment and coherence.

Despite these advantages, existing methods often fall short in achieving precise control over localized features, such as lesions, as they primarily focus on global structures. Furthermore, the generated lesions frequently lack consistency with surrounding tissues, appearing abrupt and poorly integrated. To address these limitations, we propose a hybrid mechanism based on diffusion models that combines concatenation and cross-attention for controllable mammogram synthesis. This approach ensures enhanced anatomical alignment between lesions and surrounding tissues while enabling precise control over lesion characteristics, ultimately improving both the realism and diagnostic relevance of the synthesized images.

\section{Method}

We propose the Gated Conditional Diffusion Model (GCDM), a novel framework for precise and effective mammogram synthesis that ensures accurate preservation of geometric and radiomic properties of generated masses, shown in Fig.~\ref{method}. This section first reviews the preliminaries of Denoising Diffusion Probabilistic Models (DDPM), then introduces GCDM, and finally details the gated-fusion-based Mass Control Branch.

% \begin{figure*}[htbp]
% \centering
% \includegraphics[width=\linewidth]{IJCNN25_Fig2.pdf}
% \caption{An overview of the gated fusion mechanism for precise mass control.} \label{fusion}
% \end{figure*}

\subsection{The Preliminaries of Diffusion Models}

Diffusion models~\cite{DDPM,DDIM} approximate the data distribution $ q(x_{0}) $ with a model-predicted distribution $ p_\theta(x_0) $ by reversing a noising process, comprising forward and reverse phases.

\paragraph{Forward Process}
 Data is corrupted by adding Gaussian noise with a predefined increasing normalized variance schedule $ \{\beta_t\}_{t=1}^T $ at each time step. Defining $ \bar{\alpha}_{t}=\prod_{s=1}^{t}(1-\beta_{s}) $, the noisy data $ x_t $ at step $ t $ is:
\begin{equation}
    x_t=\sqrt{\bar{\alpha}_t}x_0+\sqrt{1-\bar{\alpha}_t}\epsilon
    \label{eq:1}
\end{equation}
where $ \epsilon \sim \mathcal{N}(\textbf{0},\textbf{I}) $ and $ t \sim [1,T] $ is randomly sampled during training.

\paragraph{Reverse Process}
 A neural network $ \epsilon_{\theta} $ is trained by minimizing:
\begin{equation}
    \mathcal{L}=\mathbb {E}_{x_0,\epsilon,t} \left \|\epsilon -\epsilon _{\theta} \left ( x_t,t\right )   \right \|^{2}
    \label{eq:2}
\end{equation}

During inference, a Gaussian noise map $ x_T \sim \mathcal{N}(\textbf{0},\textbf{I}) $ is first sampled and then denoised iteratively over $ T $ steps:
\begin{equation}
    x_{t-1} = \frac{1}{\sqrt{\alpha_t}}\left( x_t - \frac{1-\alpha_t}{\sqrt{1-\bar{\alpha}_t}} \epsilon _{\theta} \left ( x_t,t\right ) \right) + \sigma_t\epsilon
    \label{eq:3}
\end{equation}
where $ \sigma_t $ depends on $ \beta_t $.

By incorporating additional conditional information $ c $ (e.g., text, labels, or images), the denoising model becomes $ \epsilon _{\theta} \left ( x_t,t,c\right ) $, ensuring the generated outputs adhere to the provided conditions.

\subsection{Holistic Mammogram Synthesis for Capturing Tissue Interaction}
Fig.~\ref{method}(right) illustrates the mammogram synthesis pipeline. Given a real clean mammogram image \( x_0 \in \mathbb{R}^{3 \times H \times W} \), where \( H \) and \( W \) denote the image height and width, respectively, we employ an off-the-shelf Variational Autoencoder (VAE) ~\cite{vae} to encode it into the latent space as \( z_0 = \mathcal{E}(x_0) \in \mathbb{R}^{4 \times 32 \times 32} \), where \( \mathcal{E} \) represents the VAE encoder. Noise \( \epsilon \sim \mathcal{N}(\textbf{0}, \textbf{I}) \) is then added to \( z_0 \) over \( t \) steps according to the forward process of diffusion models, as described in Eq.~(\ref{eq:1}), resulting in the noisy latent representation \( z_t \in \mathbb{R}^{4 \times 32 \times 32} \):
\begin{equation}
    z_t = \sqrt{\bar{\alpha}_t} \mathcal{E}(x_0) + \sqrt{1 - \bar{\alpha}_t} \epsilon
\end{equation}

During training, we utilize a denoising model \( \epsilon_\theta \) to predict the added noise, as shown in Eq.~\ref{eq:2}. In addition to the noised latent image \( z_t \), we also input an encoded three-channel binary mask \( M \in \mathbb{R}^{3 \times H \times W} \), where each channel corresponds to the background, breast tissue, and lesion (i.e., mass) regions, respectively. This mask, which includes the location and shape of both the breast and the mass, serves as a conditioning input to control both the global geometric properties of the breast and the local characteristics of the lesion.

To enhance the model's ability to capture the interaction between the breast and the lesion, we apply a Gaussian blur operation \( \mathcal{G} \) to the lesion channel of \( M \), creating a soft boundary that facilitates better integration of the mass with the surrounding tissue. The training objective is formulated as:
\begin{equation}
    \mathcal{L} = \mathbb{E}_{x_0, \epsilon, t} \left\| \epsilon - \epsilon_\theta(\mathtt{concat}(z_t, \mathcal{E}(\mathcal{G}(M))), t) \right\|^2
\end{equation}
where \( \mathtt{concat} \) denotes the concatenation operation along the channel dimension, and \( \mathcal{G} \) represents the Gaussian blur operation applied to the lesion channel.

\subsection{Gated-Fusion-Based Lesion Control Branch for Emphasizing Clinical Knowledge}

While concatenating a soft mask with the input enables rough mammogram synthesis and captures interactions between regions, it fails to achieve fine-grained control, particularly in aligning synthesized lesions with clinical knowledge. To address this, we adopt the Stable Diffusion model as our denoising framework, leveraging its text-conditioning branch to emphasize lesion features and ensure clinical alignment. However, since Stable Diffusion is trained on natural images, it lacks inherent medical concepts like `normal tissue' and `mass', often leading to poor radiological fidelity and distorted mass regions. To overcome this, we propose a gated-fusion-based Mass Control Branch, guiding the model to generate mammograms with accurate geometric and clinically relevant radiomic properties.

Fig.~\ref{method}(left) demonstrates the mass feature extraction process. To emphasize the radiomic characteristics of the generated mass, we extract radiomic features \( f \) from the lesion channel \( M_{mass} \) and the corresponding image \( x_0 \) using PyRadiomics~\cite{Radiomics}. The feature vector \( f \in \mathbb{R}^{1 \times 67} \) includes four types of radiomic features: shape (9 dimensions), histogram (18 dimensions), Gray-Level Size Zone Matrix (GLSZM, 16 dimensions), and Gray-Level Co-occurrence Matrix (GLCM, 24 dimensions). If no mass is present, \( f \) is set to a zero vector, \textit{i.e.}, \( f = \mathbf{0} \in \mathbb{R}^{1 \times 67} \). Convolutional operations are then applied to \( f \) to produce \( f_{rad} \in \mathbb{R}^{n \times 67} \). Simultaneously, to emphasize the geometric properties of the lesion, we employ CLIP~\cite{CLIP} to embed the mass mask \( M_{mass} \) and apply convolutional operations, yielding candidate geometric features \( f_{geo} \in \mathbb{R}^{m \times 768} \).

Geometric features \( f_{geo} \) capture the spatial properties of lesions, while radiomic features \( f_{rad} \) encode their textural and statistical characteristics. Combining these features ensures that the synthesized lesions exhibit both accurate shapes and clinically meaningful textures. Although directly setting \( m = n \) and concatenating the features is straightforward, geometric and radiomic features are interdependent, and direct fusion may lead to feature incompatibility or redundancy. To address this, we design a gated feature fusion mechanism to effectively integrate the most relevant geometric and radiomic features.

The geometric features \( f_{geo} \in \mathbb{R}^{m \times 768} \) and radiomic features \( f_{rad} \in \mathbb{R}^{n \times 67} \) are cross-concatenated to generate \( m \times n \) mixed candidate features, denoted as \( f_{comb} \in \mathbb{R}^{m \times n \times (768 + 67)} \), which contain \( m \times n \) feature fusion candidates.

As illustrated in Fig.~\ref{method}(center), the Gated Fusion module transforms $f_{comb}$ into the ultimate conditioning vector c for UNet. To compute the relevance score \( \mathbf{w}_{\text{gate}} \) for each feature fusion candidate in \( f_{comb} \), reflecting their intra- and inter-feature relationships, we use average pooling (AVG) to capture intra-feature information and a multi-layer perceptron (MLP)~\cite{MLP} to model inter-feature dependencies. The final relevance score is obtained by multiplying these two components:
\begin{equation}
    \mathbf{w}_{\text{gate}} = \mathbf{AVG}(f_{comb}) \odot \mathbf{MLP}(f_{comb})
\end{equation}

We then scale the original fused features using \( \mathbf{w}_{\text{gate}} \) and select the top-$k$ fused features based on their relevance scores. The final conditional features are obtained through a convolutional operation:
\begin{equation}
    \mathbf{c} = \mathtt{Conv}(\mathtt{TopK}(f_{comb} + \mathbf{w}_{\text{gate}} \odot f_{comb}))
\end{equation}

The generated features \( \mathbf{c} \) serve as the control condition for the Mass Control Branch. These features interact with the denoising model \( \epsilon_\theta \) via a cross-attention mechanism, where they act as the key and value for control, similar to other methods~\cite{LDM,ye2023ip,zero123}. This ensures that the synthesized mammograms exhibit both geometric accuracy and clinically meaningful radiomic properties.

Finally, GCDM employs two types of conditional controls: (1) the holistic mask \( M \) in concatenation, and (2) the lesion feature \( \mathbf{c} \). The denoising network \( \epsilon_\theta \) is optimized using the following objective function:
\begin{equation}
    \mathcal{L} = \mathbb{E}_{x_0, t, \epsilon} \left\| \epsilon - \epsilon_{\theta} \left( \mathtt{concat}(z_t, \mathcal{E}(\mathcal{G}(M))), t, \mathbf{c} \right) \right\|^{2}
\end{equation}

Once the denoising model is trained, we initialize the process with a random Gaussian noise map \( z_T \sim \mathcal{N}(\textbf{0}, \textbf{I}) \). The denoising model iteratively refines \( z_T \) over \( T \) steps to estimate the clean latent representation \( \hat{z}_0 \) using \( \epsilon_{\theta} \left( \mathtt{concat}(z_t, \mathcal{E}(\mathcal{G}(M))), t, \mathbf{c} \right) \) and Eq.~(\ref{eq:3}). At last, \( \hat{z}_0 \) is decoded into the synthesized mammogram image using the VAE decoder \( \mathcal{D} \) as \( \hat{x}_0 = \mathcal{D}(\hat{z}_0) \). This process not only ensures the generation of realistic mammograms with precise control over the semantic mask, which includes the lesion, breast, and their transitional areas, effectively modeling their anatomical relationships, but also emphasizes the lesion region through its morphological and radiomic characteristics, aligning it with clinical knowledge.

\section{Experiments}
\subsection{Materials and Details}
\paragraph{Dataset.}
All experiments have been conducted on the public dataset VinDr-Mammo~\cite{dataset}, using only the cranial-caudal oblique (CC) images. We split the dataset into a training set of 6704 images, a validation set of 748 images, and a test set of 1871 images, with the number of images containing mass (one or more) being 373, 45, and 104, respectively.
For all images, we first apply the algorithm for the maximum connected component to remove the extraneous black background and resize the images to $256\times 256$. Subsequently, truncation normalization is utilized to enhance the image contrast.
For images containing mass, we use the bounding-box annotations provided by the dataset as the prompt for MedSAM to obtain the instance mask. We then extract the radiomic features $f$ of the mass using PyRadiomics. Each dimension of $f$ is normalized using Min-Max normalization across the entire training dataset, and the corresponding values are applied to the validation and test set.

\begin{figure}[htbp]
\centering
\includegraphics[width=\linewidth]{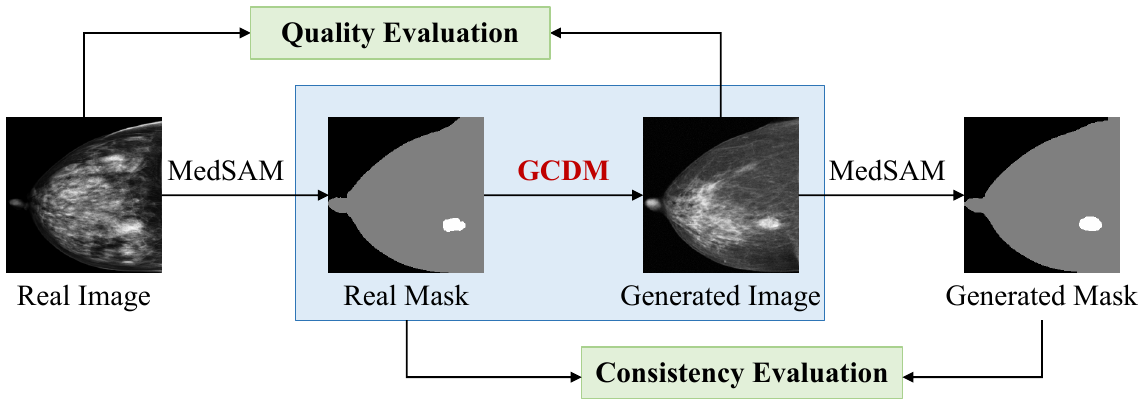}
\caption{The flowchart of quantitative evaluation. We conduct a quality evaluation between the real image and the generated image, and perform a consistency evaluation between the real mask and the generated image's mask.} \label{evaluation}
\end{figure}

\paragraph{Evaluation Metrics.}
As shown in Fig.~\ref{evaluation}, we primarily evaluate the quality of generation and the consistency learned.
For quality evaluation, we calculate the widely adopted Fréchet Inception Distance (FID), which measures the distance between the distributions of real and generated images in feature space, where features are extracted by Inception network\cite{inception}:
\begin{equation}
\text{FID} = \| \mu_r - \mu_g \|^2 + \text{Tr}(\Sigma_r + \Sigma_g - 2(\Sigma_r \Sigma_g)^{1/2})
\end{equation}
where Tr represents the trace, \( \mu_r \) and \( \Sigma_r \) are the mean vector and covariance matrix of the real images' features, and \( \mu_g \) and \( \Sigma_g \) are the mean vector and covariance matrix of the generated images' features. In the experiments, FID is calculated between all real and generated images in the test set.

The learned consistency focus on the correspondence between the generated image and the mask condition. For consistency evaluation, we use Intersection over Union (IoU) and Pixel Accuracy (PA) during real mask and corresponding generated image's mask, as shown in Fig.~\ref{evaluation}. IoU measures the overlap between corresponding regions of two masks, while PA quantifies the percentage of same pixels in the two masks. IoU and PA can be formulated as: 
\begin{equation}
    \text{IoU} = \frac{|P \cap G|}{|P \cup G|}
\end{equation}
\begin{equation}
    \text{PA} 
    % =\frac{\text{Number of correct pixels}}{\text{Total number of pixels} } 
    = \frac{\sum_{i} \mathbb{I}(p_i = g_i)}{N}
\end{equation}
where $G$ and $g_i$ represent the real mask and its pixel value, and $P$ and $p_i$ denote the generated image's mask and its pixel value. $N$ is the total number of pixels in the mask. IoU and PA are computed for each individual image in the test set and then averaged.

\begin{figure*}[htbp]
\centering
\includegraphics[width=\linewidth]{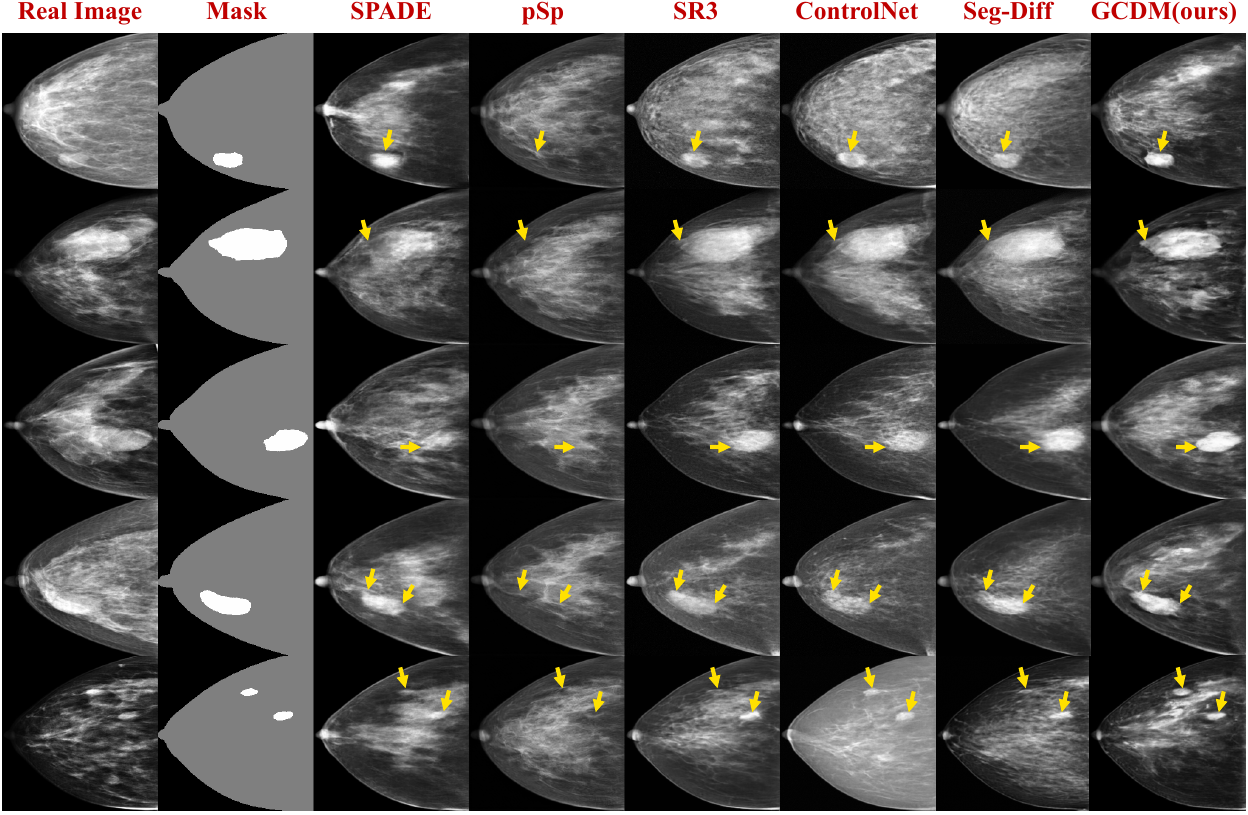}
\caption{Visual results on the VinDr-Mammo dataset. We compare our GCDM with several SOTA methods, \textit{i.e.}, SPADE, pSp, SR3, ControlNet and Seg-Diff. Our approach provides superior visual quality in terms of realism, lesion consistency, and anatomical coherence. The yellow arrows indicate the synthesized mass regions with different methods.} \label{visual}
\end{figure*}

\paragraph{Implementation Details.}
The entire framework is implemented in PyTorch, and trained on two NVIDIA A100 80GB GPUs. 
The version of Stable Diffusion we use is v1.5. In our gate fusion module, $m$, $n$, and $k$ are set to $5$.
When applying Gaussian blur to the breast mask $\mathbf{M}$, the $\sigma$ is set to $1.5$.
AdamW optimizer with an initial learning rate of $1e-4$ is used for optimization. To ensure training stability and generation equality, we set $\{\beta_i\}^T_{i=1}$ to linearly vary from $8.5e-4$ to $0.012$. 
GCDM is trained for $400$ epochs using a batch size of $32$ for each GPU.
The total steps $T$ are set as $1000$ and $50$ steps are used during inference.
We employ the CFG strategy, where during training, the input conditions are randomly masked with a probability of 0.1, and during inference, the guidance scale is set to 7.5 to achieve high-quality generation results.

\subsection{Comparison with SOTA Methods}\label{AA}

\begin{table}[htbp]
    \centering
    \renewcommand\arraystretch{1.3}
    \caption{Quantitative comparison with existing methods on VinDr-Mammo dataset. \textbf{Bold}: Best result. \underline{Underline}: Second-best result.}
    \label{tab1}
    \resizebox{\columnwidth}{!}{  % Let the table scale to fit the column width
    \begin{tabular}{c|cccc}
        \hline
        Method & FID↓ & Mass IoU \%↑ & Breast IoU \%↑ & PA \%↑ \\ 
        \hline
        SPADE & 51.60 & 83.57 & \textbf{97.67} & \underline{98.40} \\
        pSp & 45.92 & 71.51 & 95.08 & 96.65  \\ 
        SR3 & 39.28 & \underline{84.02} & 97.08 & 98.04  \\ 
        ControlNet & 33.67 & 79.36 & 97.42 & 98.27  \\ 
        Seg-Diff & \underline{30.50} & 79.34 & 96.94 & 98.01  \\ \hline
        \textbf{GCDM(Ours)} & \textbf{26.77}  & \textbf{86.30} & \underline{97.63} & \textbf{98.41} \\ 
        \hline
    \end{tabular}
    }
    
\end{table}
\begin{figure*}[htbp]
\centering
\includegraphics[width=\linewidth]{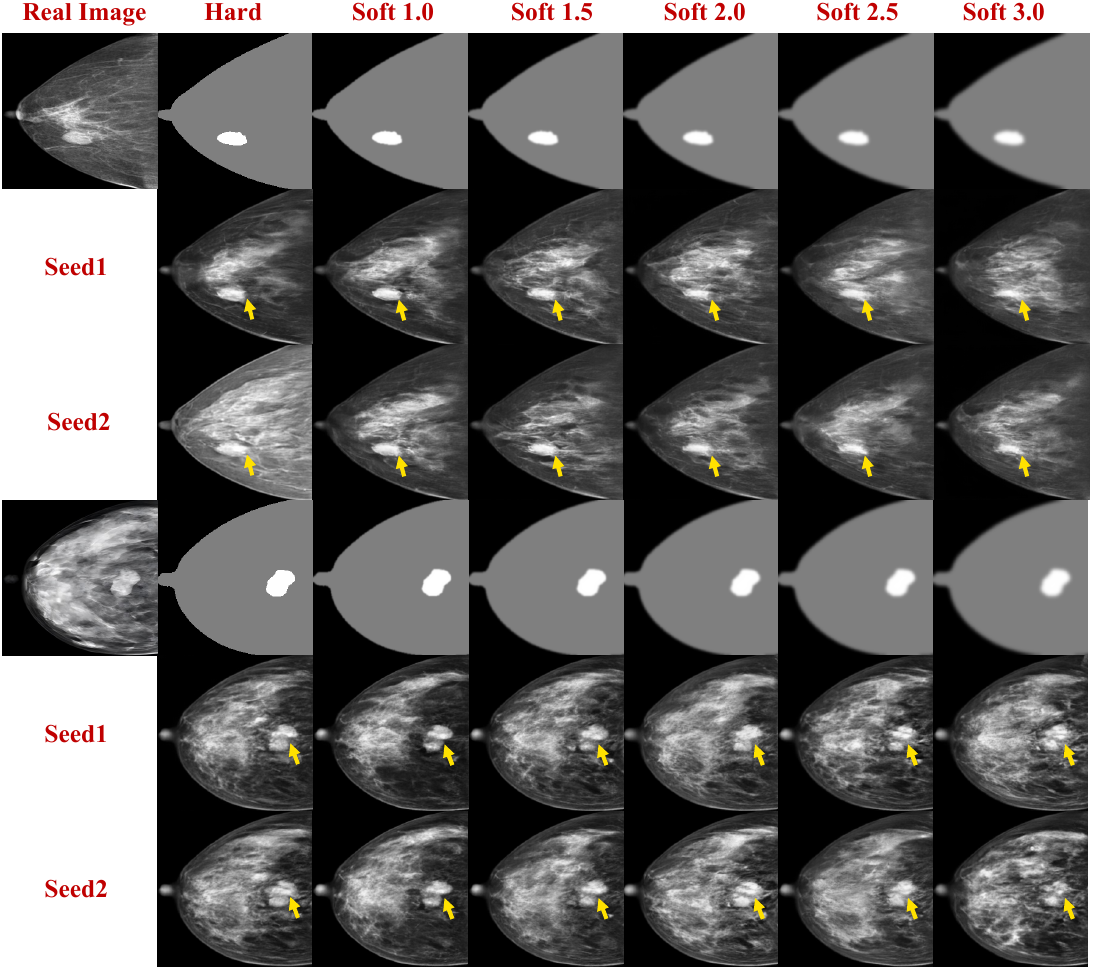}
\caption{Visualization of breast mask softness variations. Each column demonstrates progressively increasing degrees of softness, with yellow arrows highlighting the synthesized mass regions. } \label{soft}
\end{figure*}
\paragraph{Quantitative Comparison.}
We compare our GCDM with several SOTA methods on the VinDr-Mammo dataset, including GAN-based methods (SPADE~\cite{SPADE}, pSp~\cite{pSp}) and diffusion-based methods (SR3~\cite{SR3}, Seg-Diff~\cite{Segguidediff}, ControlNet~\cite{Controlnet}). All compared methods use the same train-test split as our approach.

SPADE~\cite{SPADE} and pSp~\cite{pSp} are mask-controlled synthesis methods designed for natural scenes. We utilized the official implementations to train these models for mammogram synthesis. SR3~\cite{SR3}, originally a super-resolution method, uses interpolated low-resolution images as control conditions. In our adaptation for mammogram synthesis, we replace the low-resolution images with masks as the conditioning input. For ControlNet~\cite{Controlnet}, we first fine-tune Stable Diffusion v1.5 on the mammogram dataset using the Diffusers library, then train the mask-controlled branch. For Seg-Diff~\cite{Segguidediff}, we follow the official implementation, replacing breast MRI with mammogram images.

The results shown in Table~\ref{tab1} demonstrate significant improvement in most evaluation metrics compared to the SOTA methods, particularly in the FID and Mass IoU.
Our method markedly improves FID and Mass IoU by 12.2\% and 2.71\% compared to second-best result, with corresponding p-values of 0.025 and 0.039, indicating a statistically significant improvements of our model.
The comparison results on Breast IoU and PA show that all methods perform well in global control, with our method showing only a slight improvement in PA.
Based on the results from the four metrics, our method, while achieving better performance in overall breast synthesis control as other SOTA methods, leverages our designed mass control branch to accurately and realistically generate masses that adhere to both geometric and radiomic features. This is of significant clinical importance.

\paragraph{Qualitative Comparison.}
We visually compare the synthesis results inferred by GCDM and several SOTA methods, including SPADE~\cite{SPADE}, pSp~\cite{pSp},SR3~\cite{SR3}, Seg-Diff~\cite{Segguidediff} and ControlNet~\cite{Controlnet}, on the VinDr-Mammo dataset.
Fig.~\ref{visual} presents several cases of mammogram synthesis. SPADE exhibits lower breast tissue realism, while pSp lacks control over the lesions. SR3 struggles with synthesizing multiple lesions, and both ControlNet and Seg-Diff demonstrate poor consistency between the generated lesions and the input mask. In contrast, our method synthesizes mammograms that are visually closer to real images, with richer breast tissue details. The mass produced by our method align more accurately with the semantic segmentation mask, and the transition between lesions and surrounding tissue is more natural, avoiding abrupt boundaries. Overall, our approach provides superior visual quality in terms of realism, lesion consistency, and anatomical coherence.

\subsection{Ablation Study}
In the ablation study, we first investigate the key components of GCDM, \textit{i.e.}, Lesion Condition Branch (LCB), and the subcomponents Radiomics Features (RF) and Gated Fusion (GF) within Branch. The effects of these modules are validated by sequentially adding them to the baseline, with the results shown in Table~\ref{tab2}:

\begin{table}
\centering
\caption{Ablation study of the proposed components of GCDM on VinDr-Mammo. LCB, RF, and GF represent Lesion Control Branch, Radiomics Features, and Gated Fusion, respectively.}
\label{tab2}
\resizebox{\columnwidth}{!}{
\begin{tblr}{
  cells = {c},
  vline{4} = {-}{},
  hline{1-2,6} = {-}{},
}
LCB & RF & GF & FID↓             & Mass~IoU~\%↑    & Breast~IoU~\%↑  & PA~\%↑          \\ 
$\times$ & —& —                & 31.25 &  83.48 & 96.45 & 97.62   \\
$\checkmark$ & $\times$ & —          & 29.60 &	84.97 &	96.69 & 97.98 \\
$\checkmark$ & $\checkmark $& $\times$ & 28.21 &	85.78 &	97.37 & 98.24 \\
$\checkmark$ & $\checkmark$ & $\checkmark$  & \textbf{26.77}  & \textbf{86.30}   & \textbf{97.63} & \textbf{98.41}          
\end{tblr}
}
\end{table}

\begin{itemize}
\item \textbf{Baseline}: Without the lesion control branch, the breast mask $M$ and noisy latent are concatenated as input to the denoising network $\epsilon_\theta$. This configuration results in low-quality outputs, with the highest FID value of 31.25 and a poor Mass IoU of 83.48\%.
\item \textbf{Baseline+LCB}: Adding the Lesion Control Branch, where the geometric features of the mass mask $M_{mass}$ are used as conditional controls, results in a significant improvement. The FID decreases by 1.65, and the Mass IoU increases by 1.49\%, indicating better quality and more accurate lesion control.
\item \textbf{Baseline+LCB+RF}: By adding Radiomics Features, which are concatenated with geometric features, the FID decreases by 1.39, and the Mass IoU increases by 0.81\%. While the combined condition results in more precise mass control, directly concatenating features from two different domains (geometric and radiomic) fails to effectively provide sufficient control for mass generation.
\item \textbf{Full GCDM}: When all components are included, the model achieves the best performance with an FID of 26.77, Mass IoU of 86.30\%, Breast IoU of 97.63\%, and PA of 98.41\%. Ultimately, our GCDM significantly improves the anatomical consistency, lesion control, and overall realism of the generated mammogram images.
\end{itemize}

Overall, the ablation study demonstrates the crucial contribution of the Condition Branch, including Radiomics Features and Gate Fusion, to maintaining high-quality and controllable lesion synthesis. 

\subsection{The softness of the breast mask}
In our method, we use soft labels obtained by applying Gaussian blurring to hard labels to enhance the model’s ability to capture the interaction between the breast and the lesion. Here, we investigate the impact of both hard and soft labels with varying Gaussian variances on generation performance. Considering that soft labels enhance anatomical coherence between lesions and surrounding tissues through appropriate boundary blurring, we focus on evaluating two key metrics: FID for generation quality and Mass IoU for lesion control.
The Table~\ref{tab3} presents the results for hard labels, soft labels with different variance.

\begin{table}[htbp]
\renewcommand\arraystretch{1.4}
\caption{The impact of hard labels and soft labels.}
\label{tab3}
\centering
\resizebox{\columnwidth}{!}{
\begin{tabular}{c|c|ccccc} 
\hline
Labels       & Hard  & Soft 1.0 & Soft 1.5 & Soft 2.0 & Soft 2.5 & Soft 3.0  \\ 
\hline
FID↓         & 27.23 & 28.84    & 26.77    & 22.86    & 20.62    & 20.89     \\
Mass~IoU \%↑ & 86.19 & 86.83    & 86.30    & 85.36    & 84.20    & 83.00     \\
\hline
\end{tabular}
}
\end{table}

From the results, we observe that using soft labels with a variance of 1.5 reduces the FID score and increases the mass IoU compared to hard labels, indicating improvements in both the quality of mammogram generation and the control over mass regions. As the variance of the soft labels increases further, the FID continues to decrease, suggesting enhanced image quality. However, the mass IoU drops significantly, indicating that the increased blur weakens the control over the mass boundary, as shown in Fig.~\ref{soft}. This trade-off underscores the balance between image quality and precise mass control, leading us to select soft labels with a variance of 1.5.

\section{Conclusion}

In this paper, we introduce a Gated Conditional Diffusion Model (GCDM) for controlled mammogram image synthesis, addressing the challenges of fine-grained control over lesions and maintaining anatomical consistency. Our approach incorporates a hybrid mechanism combining concatenation and cross-attention techniques within the diffusion process. The concatenation with soft labels ensures the consistency between the lesion and the surrounding tissue. And the cross-attention mechanism enables precise control over the lesion's characteristics with the gated conditioning branch, dynamically selecting and fusing the most relevant radiomic and geometric lesion properties.
Extensive experimental results on the VinDr-Mammo dataset demonstrate that our proposed GCDM significantly improves mammogram image synthesis quality and outperforms previous state-of-the-art methods in lesion control and anatomical consistency. Our method provides an effective solution for medical image generation, with substantial potential for advancing research and clinical applications.

%%
%% The acknowledgments section is defined using the "acks" environment
%% (and NOT an unnumbered section). This ensures the proper
%% identification of the section in the article metadata, and the
%% consistent spelling of the heading.
\begin{acks}
This work was supported in part by National Natural Science Foundation of China (Grant No.62202189), research grants from Wuhan United Imaging Healthcare Surgical Technology Co., Ltd.
\end{acks}

%%
%% The next two lines define the bibliography style to be used, and
%% the bibliography file.
\bibliographystyle{ACM-Reference-Format}
\bibliography{ref}

\vspace{50px}
%%
%% If your work has an appendix, this is the place to put it.
\appendix

\section{The practicality of GCDM}
\begin{figure}[htbp]
\centering
\includegraphics[width=\linewidth]{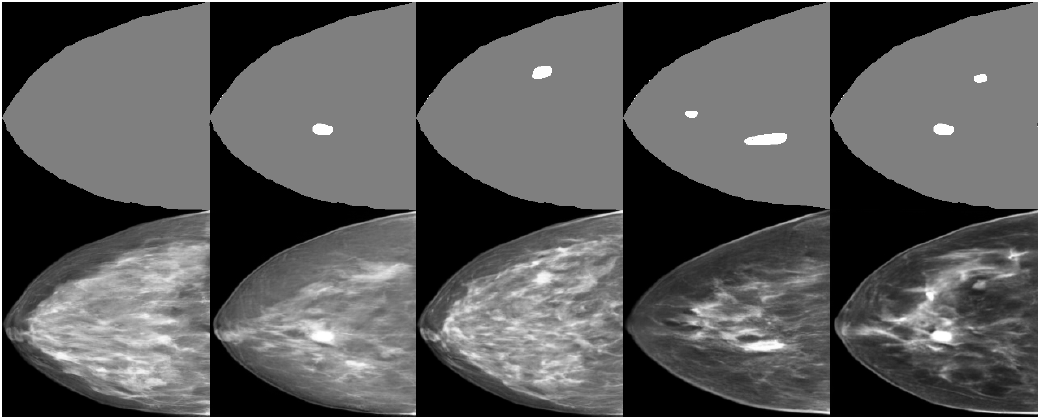}
\caption{Controllable mammogram synthesis results with GCDM. The first row shows hand-drawn masks with varying masses, accompanied by radiomic features, guiding the GCDM model to generate semantically consistent images. } \label{sample}
\end{figure}

To further evaluate the practicality of our GCDM, we also investigate the generation performance of model when only manually defined mask and radiomic features are used. Since $f$ reflects the radiomic properties of the mass and cannot be generated randomly without constraints, we first calculate the mean, minimum, and maximum value for each dimension of $f$ using all samples containing mass. 
We then define a template $f_{temp}$ using the mean values. During manual definition, for each dimension, we randomly sample a variation $\delta_i \in [\min (f_i)-f_{temp}^i, \max(f_i)-f_{temp}^i]$, resulting in the final manual radiomic features $f_{manual}$:
\begin{equation}
    f_{manual} = [f_{temp}^1+\delta_1,f_{temp}^2+\delta_2,\dots,f_{temp}^{67}+\delta_{67}]
\end{equation}
Thus, by combining the manually defined breast and mass masks, we can generate mammogram images with reasonable radiomic properties and geometric shapes.

Fig.~\ref{sample} shows mammographic images generated without a mass, with a single mass, and with multiple masses, using manually drawn breast masks and mass features generated from $f_{temp}$ as conditional controls. As observed, the generated results are highly realistic, with the interaction between the mass and surrounding tissue well. This further demonstrates the clinical applicability of our GCDM.

\section{Clinical Validation}
We conducted an additional downstream experiment to evaluate the clinical utility of our synthesized mammogram images, focusing on benign versus malignant classification. We used the VinDr-Mammo dataset and followed standard clinical practice by grouping BI-RADS categories 1–3 as benign and 4–5 as malignant. Following~\cite{ning2025retinalogos}, two widely used classification backbones were tested: ResNet-50 and ViT-B/16. Both models were trained either on real images only or on real images augmented with our synthesized data, and evaluated on a held-out real test set.

The results in Table~\ref{app:tab1} show that augmentation with our synthesized images consistently improves performance across backbones. For ResNet-50, accuracy improved from 0.673 to 0.712, AUC from 0.663 to 0.701, and F1-score from 0.605 to 0.651. For ViT-B/16, accuracy improved from 0.654 to 0.692 and AUC from 0.656 to 0.682. These findings demonstrate the practical value of our model for enhancing downstream tasks.

\begin{table}[ht]
\centering
\caption{Classification performance. Training (Tr) and test (Ts) datasets show benign/malignant case counts.}
\label{app:tab1}
\resizebox{\columnwidth}{!}{
\begin{tblr}{
  cells = {c},
  vline{2,5} = {-}{},
  hline{1-2,4,6} = {-}{},
}
Model     & Tr & Tr(ours) & Ts & Acc & AUC & F1 \\
ResNet-50 & 191/227             & -                          & 57/47           & 0.673          & 0.663          & 0.605          \\
ResNet-50 & 191/227             & 1910/2270                  & 57/47           & \textbf{0.712} & \textbf{0.701} & \textbf{0.651} \\
ViT-B/16  & 191/227             & -                          & 57/47           & 0.654          & 0.656          & 0.640          \\
ViT-B/16  & 191/227             & 1910/2270                  & 57/47           & \textbf{0.692} & \textbf{0.682} & 0.628          
\end{tblr}
}
\end{table}

\section{ Ablation w/o MedSAM}
To evaluate the impact of segmentation source on GCDM performance, we replaced MedSAM with GroundedSAM2~\cite{groundedsam} and MedSAM2~\cite{medsam2} during inference, while keeping the trained model unchanged. The synthesized images were then compared with the corresponding real images. As shown in Table~\ref{app:tab2}, different segmentation tools have limited influence on model performance, indicating that the superior results of our method primarily stem from the effectiveness of the model design itself.

\begin{table}[htbp]
    \centering
    \renewcommand\arraystretch{1.3}
    \caption{Segmentation Tools Impact on GCDM.}
    \label{app:tab2}
    \resizebox{\columnwidth}{!}{  % Let the table scale to fit the column width
    \begin{tabular}{c|cccc}
        \hline
        Tools & FID↓ & Mass IoU \%↑ & Breast IoU \%↑ & PA \%↑ \\ 
        \hline
        MedSAM       & 26.77 & 86.30        & 97.63           & 98.41   \\
        GroundedSAM2 & 26.78 & 84.75        & 98.34           & 98.88   \\
        MedSAM2      & 26.72 & 85.88        & 98.37           & 98.90   \\
        \hline
    \end{tabular}
    }
\end{table}

\section{ Different Resolutions}

We conducted experiments at both 256×256 and 512×512 resolutions by adjusting the size of the latent space in our model. Results in Table~\ref{app:tab3} show that 256×256 offers a good trade-off between performance and efficiency. Moreover, this resolution is clinically meaningful and commonly used in prior classification studies\cite{dualview,256class}. Therefore, we adopt 256×256 in the main experiments.

\begin{table}[htbp]
    \centering
    \renewcommand\arraystretch{1.3}
    \caption{Comparison of different resolutions. G denotes GPU memory usage, Tr and Ts represent training and testing, respectively, while d and s stand for days and seconds.}
    \label{app:tab3}
    \resizebox{\columnwidth}{!}{  % Let the table scale to fit the column width
    \begin{tabular}{c|cc|cccc}
        \hline
         & GPU Tr/Ts & Time Tr/Ts & FID & Mass IoU & Breast IoU & PA  \\ 
        \hline
        256                                    & 33G/17G          & 1.1d/3.2s  & 26.77 & 86.30      & 97.63        & 98.41  \\
512                                    & 77G/19G          & 2.2d/5.6s  & 27.60 & 87.07      & 97.70        & 98.46 \\
        \hline
    \end{tabular}
    }
\end{table}

\end{document}